\documentclass[aip,rsi,amsmath,amssymb,reprint,graphicx]{revtex4-1}

\usepackage{graphicx}

\begin{document}

\newcommand\EB{E_\text{B}}
\newcommand\kB{k_\mathrm{B}}
\newcommand\rw{R_\mathrm{w}}
\newcommand\rp{r_\mathrm{p}}
\newcommand\Lp{L_\mathrm{p}}
\newcommand\lambdaD{\lambda_\mathrm{D}}

\title{Plasma Temperature Measurement with a Silicon Photomultiplier (SiPM)}

\author{E. D. Hunter}
\email[]{eric.hunter@berkeley.edu}
\altaffiliation[Permanent address: ]{Stefan Meyer Institute for Subatomic Physics, Vienna, Austria}
\affiliation{Department of Physics, University of California, Berkeley, California}
\author{J. Fajans}
\email[]{joel@physics.berkeley.edu}
\affiliation{Department of Physics, University of California, Berkeley, California}
\author{N. A. Lewis}
\altaffiliation[Permanent address: ]{Department of Physics, University of Michigan, Ann Arbor, Michigan}
\affiliation{Department of Physics, University of California, Berkeley, California}
\author{A. P. Povilus}
\altaffiliation[Permanent address: ]{Lawrence Livermore National Laboratory, Livermore, California}
\affiliation{Department of Physics, University of California, Berkeley, California}
\author{C. Sierra}
\altaffiliation[Permanent address: ]{Department of Physics, University of Michigan, Ann Arbor, Michigan}
\affiliation{Department of Physics, University of California, Berkeley, California}
\author{C. So}
\altaffiliation[Permanent address: ]{TRIUMF, 4004 Wesbrook Mall, Vancouver, BC  V6T 2A3
Canada}
\affiliation{Department of Physics and Astronomy, University of Calgary, Canada}
\author{D. Zimmer}
\altaffiliation[Permanent address: ]{Department of Mechanical and Aerospace Engineering, University of California, San Diego, California}
\affiliation{Department of Physics, University of California, Berkeley, California}

\date{\today}

\begin{abstract}
The temperature of a nonneutral plasma confined in a Penning-Malmberg trap can be determined by slowly lowering one side of the trap's electrostatic axial confinement barrier; the temperature is inferred from the rate at which particles escape the trap as a function of the barrier height. In many experiments, the escaping particles are directed toward a microchannel plate (MCP), and the resulting amplified charge is collected on a phosphor screen.  The screen is used for imaging the plasma, but can also be used as a Faraday cup (FC) for a temperature measurement.  The sensitivity limit is then set by microphonic noise enhanced by the screen's high voltage bias. Alternately, a silicon photomultiplier (SiPM) can be employed to measure the charge via the light emitted from the phosphor screen. This decouples the signal from the microphonic noise and allows the temperature of colder and smaller plasmas to be measured than could be measured previously; this paper focusses on the advantages of a SiPM over a FC.
\end{abstract}

\maketitle

\section{Introduction}

Nonneutral plasmas (plasmas with a single sign of charge) can be confined in Penning-Malmberg traps. These traps \cite{malm:75}  consist of a stack of cylindrical electrodes aligned parallel to a strong axial magnetic field. The magnetic field provides radial confinement for the plasma, while potentials applied to the electrodes provide axial confinement.  The traps can confine electrons, positrons, antiprotons, ions, and single-sign mixtures, though we will here describe results with electrons only.  Penning-Malmberg traps have been used in basic plasma physics experiments on, for instance, collision rates,\cite{hyat:87,beck:92} compression,\cite{dani:05}  centrifugal separation,\cite{andr:11} cavity cooling\cite{povi:16}, electron and ion cyclotron resonance (ECR)\cite{goul:91,goul:92,sari:95,affo:14,affo:14a,affo:15} and magnetometry,\cite{hunt:20} as well as having been used to synthesize antihydrogen atoms.\cite{ahma:17}  Determining the temperature of the confined plasmas is critical to these experiments.

\begin{figure}
\includegraphics[width=\linewidth]{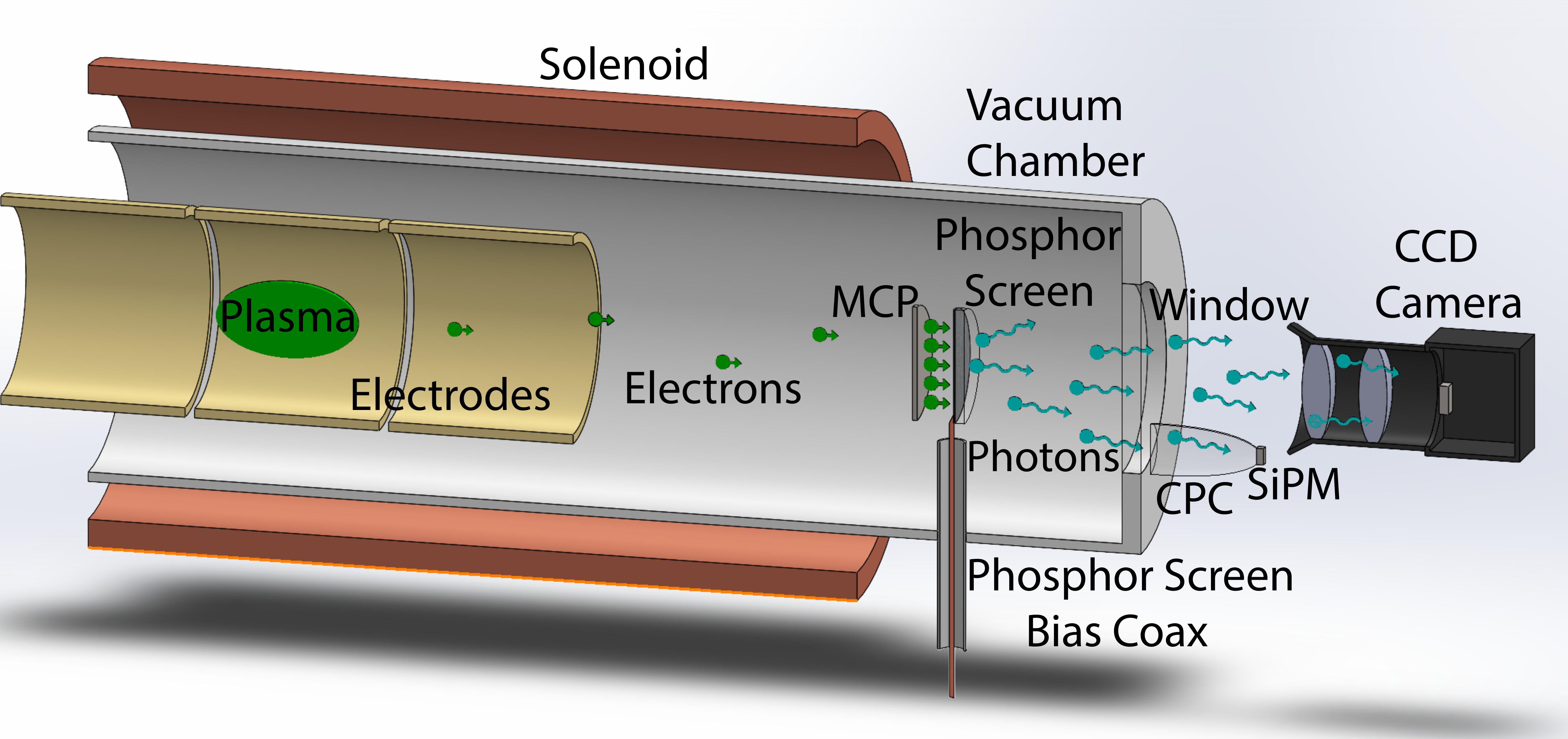}
\caption{Simplified schematic (not to scale) of the experiment. Electrons released from the plasma flow onto the MCP. Each plasma electron produces a cascade of up to $18\,000$ electrons in the MCP. These electrons are accelerated onto a fast phosphor (P47) screen to produce blue photons. The plasma signal can be read via the light collected by the compound parabolic light concentrator (CPC) and detected by the SiPM, or by the charge on the phosphor screen (FC). The CCD camera is used to determine the plasma radial density profile.}
\label{fig:experiment}
\end{figure}

\begin{figure*}
\includegraphics[width=\linewidth]{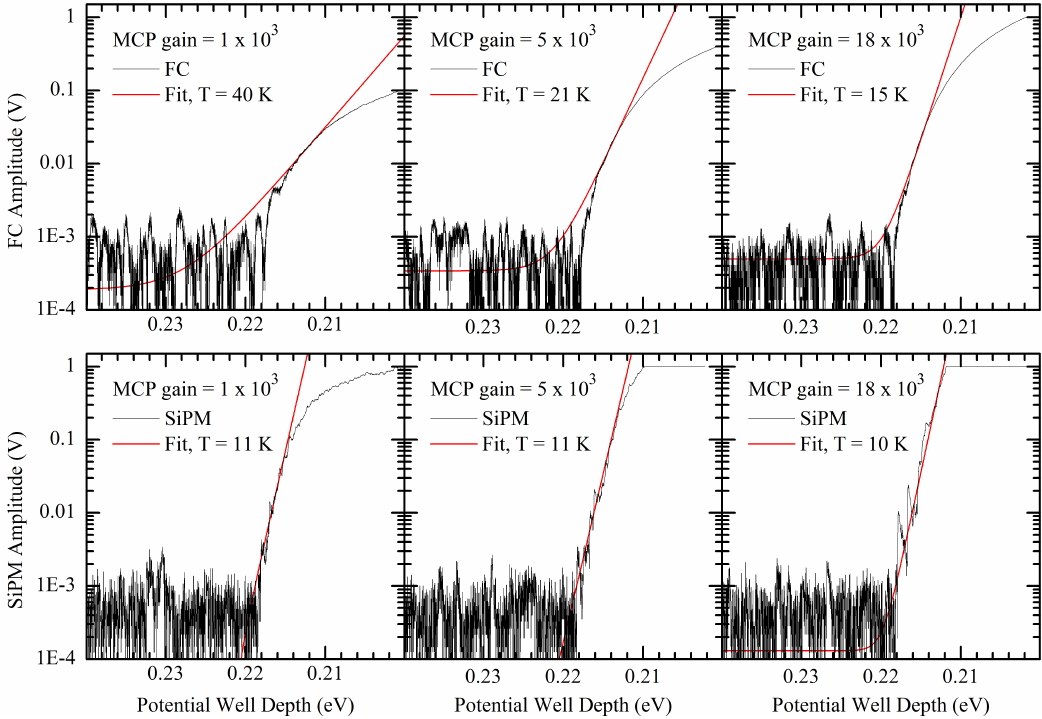}
\caption{Typical FC (top) and SiPM (bottom) extraction traces (black) and temperature fits (red) for a $3\times10^6$ electron plasma. Three MCP gain settings are tested (see legends). The absolute value is taken for both the FC and the SiPM signals so that no noise is hidden by the log scale. For all MCP gain settings, the SiPM yields significantly more linear temperature data than the FC.  The data was acquired with a sample period of $0.2\,\mu\mathrm{s}$ (FC) and $0.1\,\mu\mathrm{s}$ (SiPM); the difference does not affect the results. The ramp rate was $1\,\mathrm{V}/\mathrm{ms}$, and a $40\,\mu\mathrm{s}$ window is shown. The zero baseline of each sample was individually adjusted using the $1\,\mathrm{ms}$ of data that immediately preceded the displayed data.  Note that the ADC used to digitize the signal saturates at $1\,\mathrm{V}$ and sets the maximum value observed for all traces.}
\label{fig:sidebyside}
\end{figure*}

For electrons and many ions, fluorescent-based temperature diagnostics are not available.  Instead, the most common temperature diagnostic functions by measuring the rate that electrons ``evaporate'' as one of the side well barriers is lowered (see Fig.~\ref{fig:experiment}).\cite{eggl:92} More precisely, we obtain $N(\EB)$, the often microchannel-plate-(MCP)-amplified number of electrons that escape the plasma as a function of the energy barrier height $\EB(t,r)$
\begin{equation}
\label{eq:barrier_height}
\EB(t,r)=-q[V(t,r)-\Phi(t,r)],
\end{equation}
where $q$ is the unit charge, $r$ is the radius from the trap axis, $V(t,r)$ is the depth of the vacuum electrostatic well created by the time ($t$)-dependent voltages applied to the confining electrodes, and $\Phi(t,r)$ is the plasma's time-dependent self-consistent potential. The temperature $T$ is then determined by fitting the rising edge of the signal, $N(\EB)$, to an exponential of the form
\begin{equation}
\label{eq:lin_fit}
N(\EB)\propto\exp[-\EB(t,r)/\kB T],
\end{equation}
where $\kB$ is Boltzmann's constant.\cite{hyat:88,beck:90,eggl:92,beck:96}  Typical examples of $N(\EB)$ and the resulting fits are shown in Fig.~\ref{fig:sidebyside}.  One can see in the figure, particularly in the Faraday Cup (FC) examples, that the exponential region exists, at best, only at low signal amplitudes.

In principle, the temperature $T$ could vary as a function of the plasma radius $\rp$.  If this occurred it would invalidate the simple exponential model in Eq.~(\ref{eq:lin_fit}), and require us to resort to more advanced methods.\cite{eggl:92}  However, for cold lepton plasmas, which are the concern of this paper, the kinetic energy distribution of the electrons is well-thermalized. With typical plasma radii of the order of several millimeters, the thermal relaxation time in the relevant ``long range'' regime is on the order of milliseconds,\cite{dubi:97,holl:00a} and the plasmas generally have ample time to relax before their temperature is measured.

The exponential form of $N(\EB)$ results from the assumed Maxwellian distribution of the plasma electrons, and is only valid for electrons that come from the high-energy tail of the distribution. The particles near the axis face a lower energy barrier, and escape first.  As ever more electrons escape, the self-consistent plasma potential $\Phi(t,r)$ begins to change.  The self-consistent barrier height $\EB$ becomes increasingly dependent on $N(\EB)$, and $N(\EB)$ itself becomes increasingly independent of the temperature.  Thus, only electrons that escape from the inner radial core of the plasma contain temperature information that strictly follows Eq.~(\ref{eq:lin_fit}).\cite{beck:90,eggl:92,beck:96}  This cylindrical core, roughly confined within a radius that is of order one Debye length\cite{evan:16a,adri_private} ($\lambdaD$), contains $\sim0.1\pi\epsilon_0 \kB T \Lp/q^2$ electrons, where $\Lp$ is the plasma length.  (For any $\EB$, the core is not uniformly extracted from the plasma; on the axis a larger fraction of the electrons will have escaped than at $r=\lambdaD$.)

Notably, the number of electrons in the core is independent of the plasma density $n$. At low temperatures, this number is not large.  For instance, for a plasma with $T=10\,\mathrm{K}$ and $\Lp=5\,\mathrm{mm}$, less than $100$ electrons contain temperature information readily fit to an exponential.  A sensitivity approaching one electron charge is necessary to accurately measure the temperatures of plasmas with such parameters.

Often, $N(\EB)$ is collected using the phosphor screen in an MCP/phosphor screen imaging diagnostic as a Faraday cup (FC). (The charge can also be collected on a regular FC, generally without a MCP for electron amplification.) The screen must be biased to a voltage higher than the accelerating voltages employed by the MCP: typically, a minimum of $\sim 1\,\mathrm{kV}$.  If the screen is simultaneously used for imaging, the bias must be increased to $4.5\,\mathrm{kV}$ or above, since the electrons require $1\mbox{--}2\,\mathrm{keV}$ to penetrate the aluminium coating on the screen.  The signal, $N(\EB)$, comes from a capacitive pickoff on the screen bias circuitry. The dominant noise on the pickup is microphonic, and at least partially generated by small changes in the  capacitance of the cable and of the pickoff capacitor as they move in response to ambient vibrations.  Since the resulting charge fluctuations are proportional to the voltage on the cable and capacitor, the microphonic noise is enhanced by the required high bias voltages.  The ambient vibrations that induce this noise can be strongly enhanced by the proximity of active cooling systems associated with the magnet and cryogenic plasma trap. Even without such cooling systems, however, microphonic noise generally dominates over electronic noise in the system.

The technique presented in this paper bypasses the bias coupling by using the blue light emitted from the P47 phosphor screen as a proxy for the incident charge. This light is unaffected by the vibrations, and, hence, is immune to the microphonic noise.  To obtain $N(\EB)$ from the phosphor light, we have used both traditional photomultiplier tubes and silicon photomultipliers (SiPM). Although photomultiplier tubes may have a lower dark count than SiPMs, they must be placed some distance away from the experiment because of the magnetic fringe field of the Penning-Malmberg trap.  Even so, with extension optics they yield a superior signal compared to the signal from a Faraday cup.

In this paper, we describe light-collection results with SiPMs only because of their magnetic field insensitivity,\cite{hawk:07} which allows them to be placed just outside the trap vacuum window, much closer to the phosphor screen.  Consequently, they can collect much more light than a photomultiplier tube, and in this application, yield equivalent results.  Moreover, unlike photomultiplier tubes, SiPMs are not delicate, are relatively inexpensive, and are not degraded by accidental exposure to ambient light.\cite{buzh:02,godf:18,stag:18}

\section{Data Collection and Temperature Fitting}
Before discussing the circuitry to collect $N(\EB)$, it is worth discussing some general aspects of the data collection process and temperature fitting.  One immediate observation is that the gain of data collection circuitry is only a nuisance parameter in fitting for $T$ in Eq.~(\ref{eq:lin_fit}).  A less obvious observation is that, so long as $\EB(t,r)$ scales linearly with time, passing the incoming data through a low or high pass filter does not change the fit; an exponential passed through such a filter remains an exponential with the same growth constant.  Thus, though the signal from the FC is actually integrated, $\int^{t}_{-\infty}d{\bar t}\,\EB(\bar{t},r)$, equivalent to passing the signal through a low pass filter, this has no effect on the temperature fitting.

In principle, we can tune the slope $d\EB(t,r)/dt$ to best employ filters to increase the signal-to-noise (SNR) ratio.  In practice, there are limitations.   The slope of $\EB(t,r)$ determines the dump time interval over which the plasma escapes following a pure exponential [Eq.~(\ref{eq:lin_fit})].  If the dump time is too short, an individual particle's transit time through the plasma can cause its escape to be delayed.  These transit times are on the order of $1\,\mu\mathrm{s}$ and set a lower limit on the dump time.  If the dump time is too long, hollowing instabilities related to diocotron\cite{levy:65,dris:90a} and Kelvin-Helmholtz\cite{peur:93b} instabilities are frequently observed.  The growth time for these instabilities scale with the plasma rotation frequency (sometimes reduced by the ratio of $(\rp/\rw)^2$, where $\rw$ is the trap wall radius), and range over $\sim 10$---$1000\,\mu\mathrm{s}$.  (Examples of these instabilities are shown later in this paper in Fig.~\ref{fig:cat}.)  These instabilities set an upper limit on the dump time.  In practice, a dump time on the order of ten microseconds often works well.

In many diagnostics, the effects of noise can be reduced by averaging multiple data sets.  In our case, we could average the results [the $N(\EB)$] of dumping multiple plasmas.  Unfortunately, averaging does not generally work for our diagnostic.  Plasma-to-plasma variations are often on the order of $1$\%, and these variations change $\Phi(r,t)$ proportionally.  The net effect is that the composite signal is an average of exponentially rising signals randomly shifted in time. While this average will still be a net signal with an appropriate rise time, the SNR will often be dominated by the earliest arriving signals and will not significantly improve.  For typical plasmas ($\rp=1\,\mathrm{mm}$, $\rw=20\,\mathrm{mm}$, $n=10^8\,\mathrm{cm}^{-3}$) with $1$\% variations, the utility of averaging will fade for temperatures below several hundred kelvin. Above this temperature limit, the SNR of a single signal is often adequate and averaging would not be necessary.

\section{Detector Circuitry and Sensitivity}
\subsubsection{Faraday Cup Circuit}

\begin{figure}
\includegraphics[width=\linewidth]{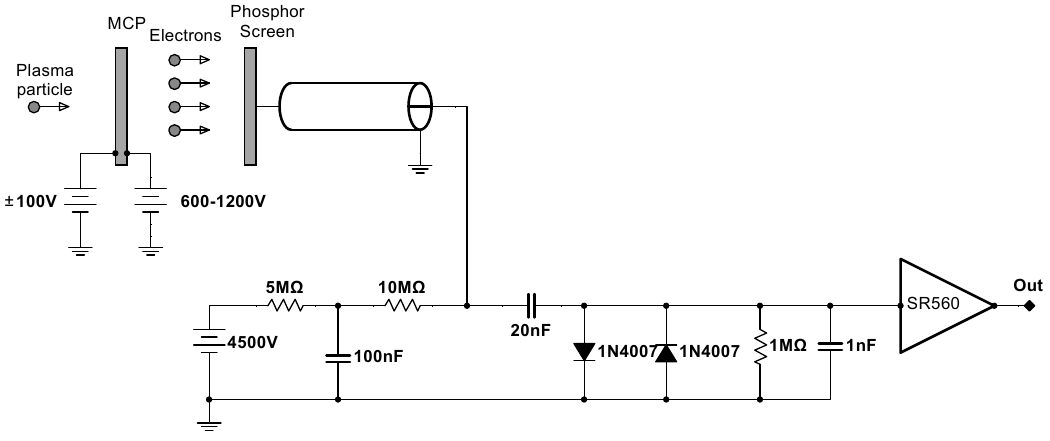}
\caption{Biasing circuit and AC coupled pickoff for the phosphor screen FC. In addition to the capacitors explicitly shown in this schematic, there is also $\sim 300\,\mathrm{pF}$ of ``parasitic'' capacitance on the coax lead going to the screen.}
\label{fig:biaschain}
\end{figure}

Our FC-based diagnostic measures the plasma charge using the circuit in Fig.~\ref{fig:biaschain}. The RC network on the right filters noise from the high-voltage power supply. The $20\,\mathrm{nF}$ pickoff capacitor blocks the DC bias voltage on the phosphor screen while coupling the signal to the subsequent SRS SR560 $1\,\mathrm{MHz}$ low pass amplifier.  The back-to-back diodes protect the SRS amplifier from voltage spikes when the high voltage supply is turned on and from faults.  One plasma electron impinging on the MCP multiplies to $\sim 18\,000$ electrons at maximum MCP gain.  These electrons are accelerated towards, and then collected on, the phosphor screen, where they induce a $\sim 2.2\,\mu\mathrm{V}$ signal on the $1\,\mathrm{nF}$ output capacitor. (The total system capacitance, including the cables and other parasitic capacitors, is about $1.3\,\mathrm{nF}$.) This charge decays through the $1\,\mathrm{M}\Omega$ resistor.

\begin{figure}
\includegraphics[width=\linewidth]{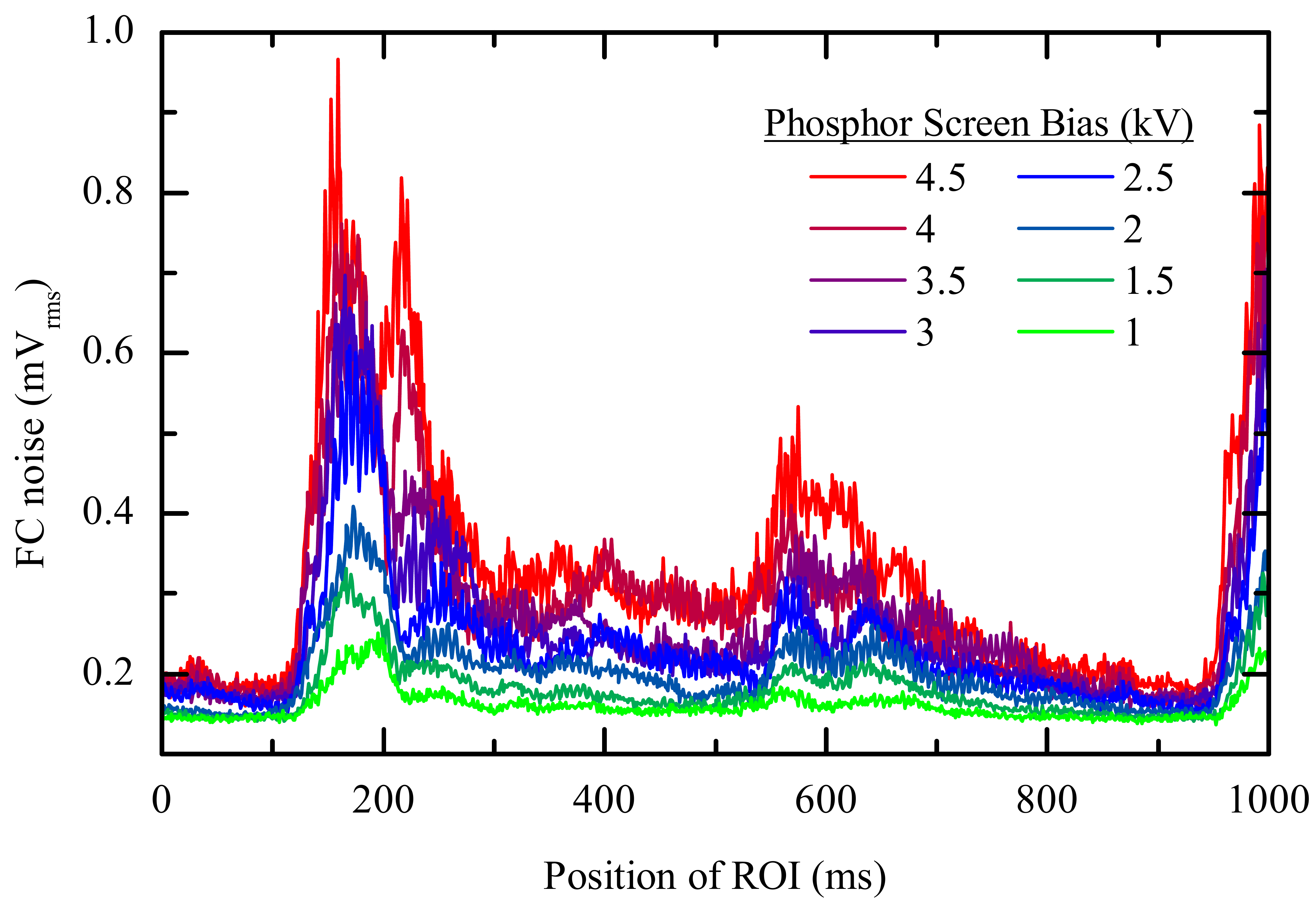}
\caption{FC noise due to microphonics related to the coldhead, and measured on the $1\,\mathrm{nF}$ capacitor in the circuit of Fig.~\ref{fig:biaschain} as a function of the phosphor screen bias. The RMS noise (minus DC offset) is measured for a $1\,\mathrm{ms}$ moving window (ROI), and averaged over $30$ traces. Traces were collected with a trigger synchronized to the coldhead cycle.}
\label{fig:FCnoise}
\end{figure}

The microphonic noise caused by vibrations of the high-voltage phosphor screen bias lead and decoupling capacitor is the primary source of noise.  The noise in our particular system is dominated by vibrations from the coldhead (Sumitomo RDK-415D) that cools the electrode stack. The coldhead generates periodic noise associated with intervals of maximum and minimum vibrations. To study this effect, we created an audio trigger synchronized with the loudest part of the periodic coldhead cycle. Figure~\ref{fig:FCnoise} displays the RMS noise in the FC signal as a function of time from this trigger. The period of elevated noise, beginning around $100\,\mathrm{ms}$, persists at some level for nearly the whole trace. The noise increases as the phosphor bias is raised toward the level required for imaging ($4.5\,\mathrm{kV}$).

Even if the plasma temperature is measured during the FC quiet intervals, which is sometimes possible, the noise is $0.1\mbox{--}0.2\,\mathrm{mV}$.  The noise principally comes from ambient vibrations, with a component from the high-voltage power supply.  For reference, the Johnson noise is about $0.002\,\mathrm{mV}$, the noise from the SRS amplifier is about $0.004\,\mathrm{mV}$, and the ADC noise is about $0.03\,\mathrm{mV}$. As with the SiPM data, the noise floor is unaffected by the MCP gain.

\subsubsection{Faraday Cup Calibration and Signal-to-Noise Ratio}

To optimize the SNR for FC-based measurements, we run the MCP at maximum gain: a front-to-back bias differential of about $1\,\mathrm{kV}$. The phosphor screen is normally biased a few hundred volts higher than the back of the MCP in order to attract and capture all the electrons leaving the MCP while minimizing the microphonic noise.

The noise is best characterized in terms of the number of equivalent plasma electrons.  This number can be estimated from the FC data in Fig.~\ref{fig:sidebyside}.  This data was taken with an SR560 gain of $2$, so a single plasma electron generates a signal of about $0.004\,\mathrm{mV}$.  The FC RMS noise level, taken before the signal starts to rise, is about $0.51\,\mathrm{mV}$.  Thus, the noise is equivalent to the signal from about $120$ plasma electrons.  Our circuit functions as an integrator with a time constant long compared to the observation time; we would expect to be able to distinguish a signal once the number of electrons that have escaped the plasma exceeds $120$.

As can be seen in Fig.~\ref{fig:sidebyside}, the signal initially doubles over a time, corresponding to a signal frequency, in which there is significant noise variation; i.e.\ the signal and noise frequencies overlap.  Consequently, it is not possible to greatly improve the SNR by filtering.

Across all five\cite{beck:92,hans:96,andr:11,povi:16,ahma:17} of the Penning-Malmberg traps that the authors have worked on, at three different institutions, microphonic noise has been the dominant noise source for FC pickups. The best noise performance has been approximately the same (within about an order of magnitude) on experiments that use an MCP\cite{hans:96,andr:11,povi:16,ahma:17,peur:93a,andr:09} and an order of magnitude worse on experiments using a cryogenic amplifier placed close to the FC.\cite{beck:92}  Attempts to bring a raw FC, no-MCP signal out of the vacuum chamber dramatically increase the noise.  This is particularly true for cryogenic traps, as cryogenic coax cables appear to be particularly noisy.

\subsubsection{Silicon Photomultiplier Circuit}
\begin{figure}
\includegraphics[width=\linewidth]{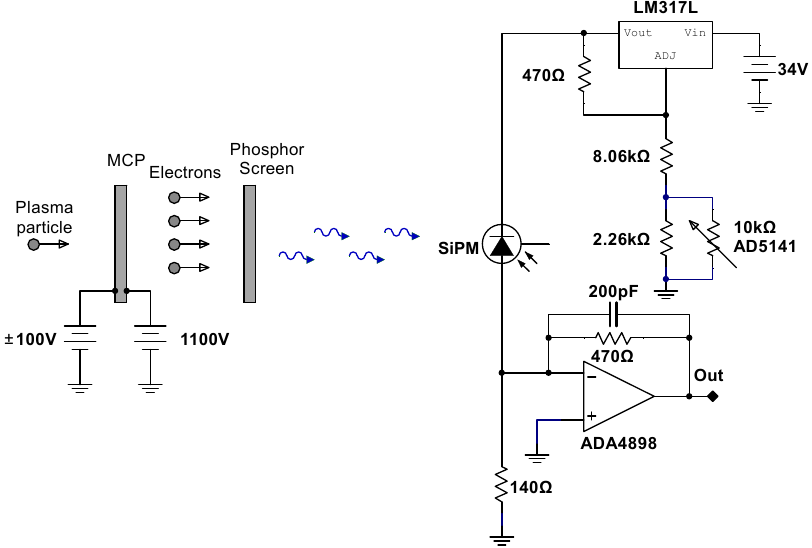}
\caption{Simplified schematic of the SiPM circuit.}
\label{fig:SiPMSchematic}
\end{figure}

We use the SensL C-Series 60035 SiPM, a solid state device constructed of $19\,600$ cells in a $6\,\mathrm{mm}$ by $6\,\mathrm{mm}$ square, though we believe similar results would be obtained for any similar SiPM. Each cell contains one Geiger-mode photodiode and a series quench resistor.  The cells themselves are in a parallel arrangement, resulting in a total output signal that is the sum of the entire array of cells. The device receives a positive reverse bias at the (parallelized) cathode, which normally exceeds the reverse breakdown voltage of $\sim 24.5\,\mathrm{V}$ by a few volts (the ``overvoltage''). The gain is a sensitive function of the overvoltage. The output is taken from the anode and amplified by an ADA4898-based transimpedance amplifier.  It is linearly related to the light intensity as long as 1) the voltage developed at the anode is much less than the overvoltage, a condition enforced by the transimpedance amplifier, and 2) the individual cells do not saturate from multiple photons. The transimpedance amplifier has a low pass $3\,\mathrm{dB}$ frequency of $1.7\,\mathrm{MHz}$.  A simplified schematic of the SiPM circuit is shown in Fig.~\ref{fig:SiPMSchematic}.

The primary noise source within the SiPM is the dark count rate (DCR): the rate at which thermal excitations trigger electron avalanches in the cells. The DCR increases with the cell temperature, the detector area, and the overvoltage. The overvoltage is the most readily adjustable parameter, but it also controls the gain of the SiPM. A variable bias voltage was incorporated into the SiPM circuit to enable precise tuning of the overvoltage to achieve optimal performance.

The bias voltage is set by changing the resistance of a non-volatile digital potentiometer (AD5141) in the feedback loop of an adjustable voltage regulator (LM317L). The AD5141 is, in turn, set over the SPI bus by an Arduino microcontroller which is ultimately disconnected from the circuit. A digital potentiometer was chosen over its manual equivalent due to the former's consistent, quiet nature and remote programmability.

The optimal bias voltage---roughly $28.5\,\mathrm{V}$ in our case---corresponds to the peak SNR. This value is determined by exposing the SiPM to light pulses from a blue LED and measuring the maximum amplitude response when normalized by the RMS noise.

To allow room for a CCD camera, which is used to image the plasma, the SiPM is placed off the trap's optical axis (see Fig.~\ref{fig:experiment}). This decreases the amount of light collected by about a factor of two. This signal loss is compensated for by the inclusion of an acrylic compound parabolic concentrator (CPC), which concentrates light onto the SiPM. The CPC has optical grease applied to both ends and silver-coated sides to maximize internal reflections. The light gain of the CPC was measured using a digital camera and is approximately a factor of $5$.  In experiments not reported here, we have also used an open-center, tilted Fresnel lens to gather even more light.

\subsubsection{Silicon Photomultiplier Calibration and Signal-to-Noise Ratio}

For SiPM-based measurements, the phosphor screen must produce light.  This requires a screen bias voltage much higher than that used for the FC-based measurements.  Typically we use the same bias as for imaging the plasma: $4.5\,\mathrm{kV}$ or greater. The SNR is highest with the MCP gain at its maximum, and we will report the SNR under these conditions.  However, we can use lower MCP gains and still measure temperatures satisfactorily; this has some advantages which will be discussed later.

\begin{figure}
    \centering
    \includegraphics[width=\linewidth]{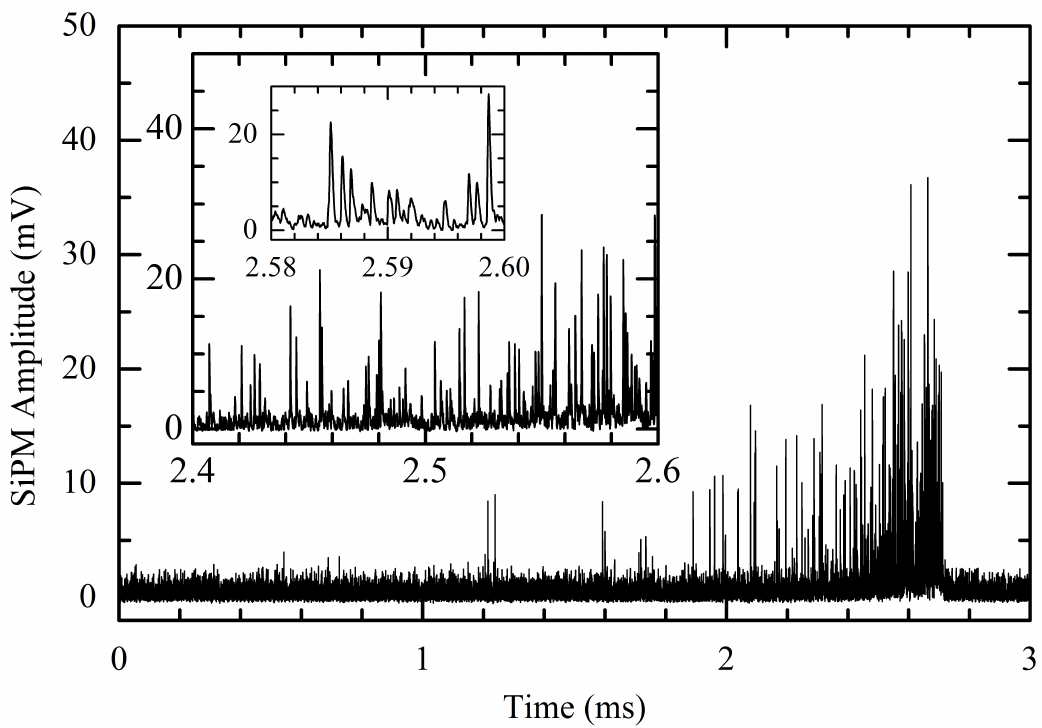}
    \caption{Typical extraction trace for $\sim 100$ trapped electrons. The larger inset shows that most hits appear to be isolated even in the densest portion of the extraction.  The smaller inset shows the time history of individual hits.}
    \label{fig:singles}
\end{figure}

To determine the mean amplitude of single plasma-electron events, we slowly release a small number of electrons ($\sim 10^2$) from the plasma onto the MCP so that each arriving electron is well separated in time (see Fig.~\ref{fig:singles}). Each ``hit'' has a sharp rise followed by an exponential decay with a time constant of about $300\,\mathrm{ns}$.

\begin{figure}
    \centering
    \includegraphics[width=\linewidth]{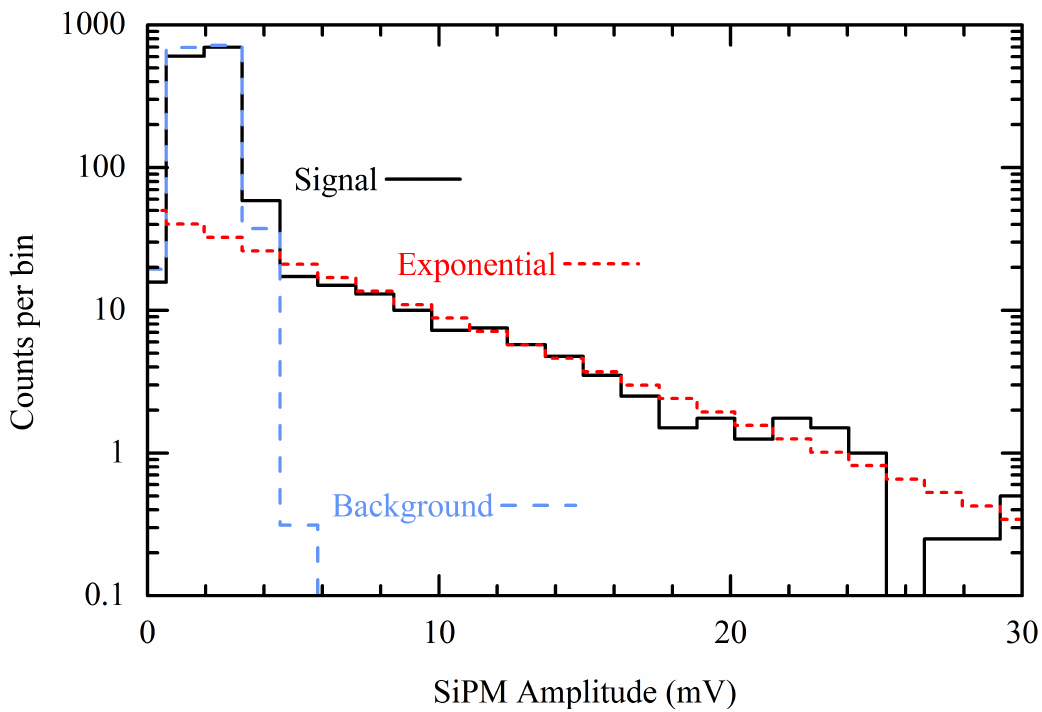}
    \caption{Averaged pulse height distribution (black) for four $\sim 100$-electron traces similar to that shown in Fig~\ref{fig:singles}. The background, without any electrons, is shown in blue, and an exponential fit is displayed in red.}
    \label{fig:PHD}
\end{figure}

Figure~\ref{fig:PHD} displays the averaged pulse amplitude histogram of four $3\,\mathrm{s}$ samples similar to that shown in Fig.~\ref{fig:singles}. The MCP gain was at its maximum, and the phosphor bias was $4.5\,\mathrm{kV}$.  In addition to the signal from the roughly $100$ electrons per sample, this histogram includes the SiPM dark counts that occurred during the samples.  With these counts subtracted, the analysis shows that the signals from the plasma electrons have a broad distribution with a mean height of about $9\,\mathrm{mV}$.

Some individual electron pulses are visible in the SiPM data in Fig.~\ref{fig:sidebyside}, particularly for the $18\times 10^3$ gain trace, where they appear as sawtooth jumps in the signal.  Once the electrons arrive on the MCP faster than a few per microsecond, the responses to individual electron overlap, and the curve appears to rise ever more smoothly.


With our preferred SiPM overvoltage, the average Poissonian interval between counts is $1\,\mu\mathrm{s}$.  This DCR is much higher than it would be if we had used a photomultiplier tube instead of a SiPM.\cite{buzh:02} Nonetheless, it causes few operational difficulties because the dark count amplitudes are low compared to the signal counts. These counts were separately measured in Fig.~\ref{fig:PHD} by averaging sixteen $3\,\mathrm{s}$ samples in which no electrons were released onto the MCP; the counts have a mean of about $2\,\mathrm{mV}$.

From Fig.~\ref{fig:sidebyside}, the calculated SiPM RMS noise level, taken before the signal starts to rise, is about $0.57\,\mathrm{mV}$.  This is as expected for $2\,\mathrm{mV}$ dark count pulses, exponentially decaying in $300\,\mathrm{ns}$, and occurring every $\sim 1\,\mu\mathrm{s}$.  Since the signal from one plasma electron is about $9\,\mathrm{mV}$, we can easily detect single electrons; the average background is equivalent to less than $0.1$ plasma electrons.

In principle, we could use a threshold to remove the dark count noise, and use the counts timestamps themselves rather than the count amplitudes to do our temperature analysis.  A $4\,\mathrm{mV}$ threshold would accept less than $0.1\%$ of the dark counts while retaining $80\%$ of the signal counts. However, we usually do not do this because pile-up of the late-released electrons would lead to missed counts and confuse the analysis.

\section{Temperature Fitting Procedure and Performance}

As discussed in the introduction, the plasma temperature is determined from the relation $N(\EB)\sim\exp[-\EB/\kB T]$, the MCP-amplified number of electrons that escape the plasma as the plasma confinement barrier is lowered.   This relation is only valid for the first $\sim 0.6 \pi\epsilon_0 \kB T \Lp/q^2$ electrons to escape, but for these first electrons, $N(\EB)$ will increase linearly on a log scale. This log-linear regime is the focus of this paper. In practice, the data is fit to a slightly extended model, $a+b\exp[-\EB/\kB T]$.  Here, $a$ is the offset from the amplification chain and from low frequency noise, and $b$ incorporates the plasma escape time, the system gain, and the plasma length; both $a$ and $b$ are nuisance parameters.

In Fig.~\ref{fig:sidebyside}, we compare the output of the FC and SiPM detectors at three different MCP gain settings.  The six measurements were not acquired simultaneously, but were made on six nominally identical plasmas.  The acquisition system saturates at $1\,\mathrm{V}$; with a 14 bit analog to digital converter (ADC), the smallest level we can measure has a magnitude of $\sim 0.1\,\mathrm{mV}$.  The noise level of both the FC and the SiPM is, coincidentally, somewhat less than $1\mathrm{mV}$.

The FC and SiPM signals in Fig.~\ref{fig:sidebyside} are qualitatively different.  The FC signal is rounded, and it is not easy to identify a linear region at any MCP gain. The SiPM signal, on the other hand, is linear for $2$--$3$ orders of magnitude.  Roundedness is only visible for the lowest MCP gain SiPM curve.  The signals are different because the SiPM is several orders of magnitude more sensitive to plasma electrons than the FC. The SiPM reaches down to much lower $N$, i.e., it detects particles that escape the plasma much earlier and originate from much closer to the plasma axis.



The SiPM signals are easy to fit to a straight line.  The fitting routine requires the identification of an upper fitting bound: the highest amplitude signal where the signal is assumed linear.  This bound can be $1\,\mathrm{V}$, the saturated signal level, but will be lower if the signal is rounded.  We find this bound with a code that minimizes the error in the fit.\cite{evan:16a}.  For the SiPM data, the routine makes very defensible choices as evidenced by the overlap between the data and the red fit line in Fig.~\ref{fig:sidebyside}.  The rounded FC signals cannot be so readily fit.  The upper fitting bound for the FC signal is not obvious by eye or by code, and the FC fits, while perhaps in the neighborhood, are obviously not precise.  The lower the assumed bound, the lower the resulting fit temperature, so one could postulate that for the FC, the fit temperatures are an upper bound.

\begin{figure}
\includegraphics[width=\linewidth]{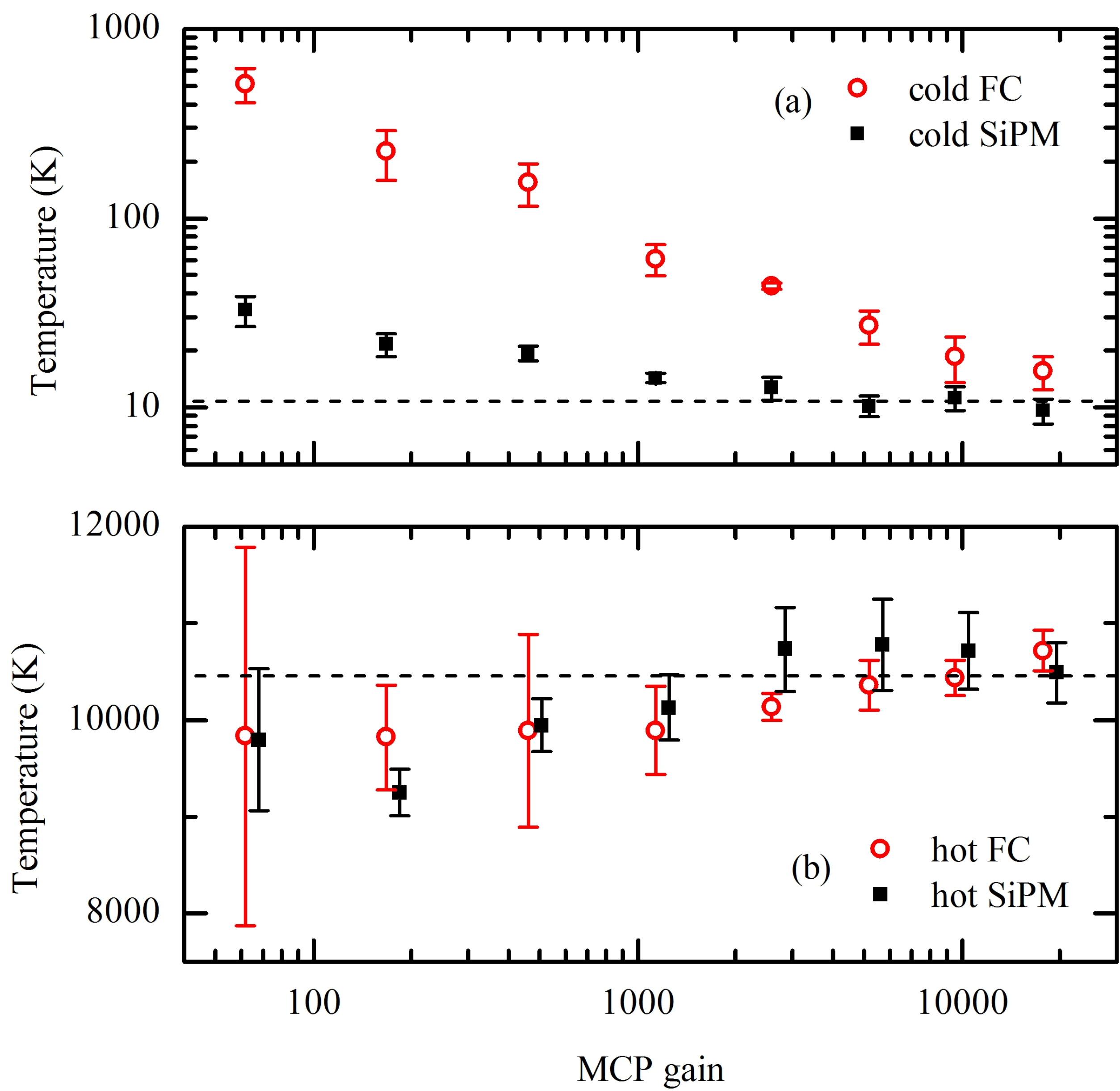}
\caption{Measured plasma temperature as a function of MCP gain for (a) very cold and (b) very hot plasmas. Error bars represent the standard deviation of the temperature measured in multiple trials; in (b), the SiPM points were horizontally displaced around their true center for visual clarity.}
\label{fig:MCPgain}
\end{figure}

In Fig.~\ref{fig:MCPgain}, we show the temperatures reported for very cold ($\sim 10\,\mathrm{K}$) and very hot ($\sim 10\,500\,\mathrm{K}$) plasmas as a function of MCP gain.  For cold plasmas, the SiPM temperatures asymptote to the temperature found with the highest MCP gain, and are within about $20\%$ for all gains above $\times 2000$. The FC temperatures asymptote towards this same value, but much more slowly.  As postulated, the FC temperatures are in all cases too high (orders of magnitude too high for low MCP gains). For hot plasmas, the measured temperatures for both detectors converge to the highest MCP gain value, and are never more than about $10\%$ from this value.

As discussed above, the temperature information in $N(\EB)$ diminishes beyond the pure exponential regime of Eq.~\ref{eq:lin_fit}.  However, it does not disappear entirely, and it is possible to fit a known universal function\cite{beck:90,beck:96} to the data well beyond the linear regime.  This allows one to use higher amplitude data without increasing the instrumental noise, thereby increasing the SNR.  However, this is not a panacea.  The hollowing instabilities will still limit the valid data collection time, and the onset of these instabilities is not always obvious.  Moreover, the universal fit employs three nuisance parameters (gain$\times \Lp$, $\rp/\lambdaD$, and $\rp/\rw$). This increases the error in fitting $T$, particularly because the high amplitude data is dominated by the nuisance parameters. Further, unlike the shape of the simpler pure exponential function, the shape of the universal function is not preserved by filtering. Finally, the universal function uses a model of the plasma and trap that is not necessarily correct.  It assumes that the temperature is uniform across the entire plasma, not just in a near-axis core.  The evaporative cooling of the early escaping particles can cool the late escaping particles.  The fit is exquisitely sensitive to the assumption that the plasma assumes the shape of a uniform density, right cylinder.  In practice, the plasmas are generally ovoids.  It also assumes that the vacuum potential increases as $r^2$; this is only approximately true.  All these considerations make the ``universal'' model difficult to use and we do not employ it here.

\section{Plasma Space Charge and Simultaneous Imaging}
\begin{figure}
    \centering
    \includegraphics[width=\linewidth]{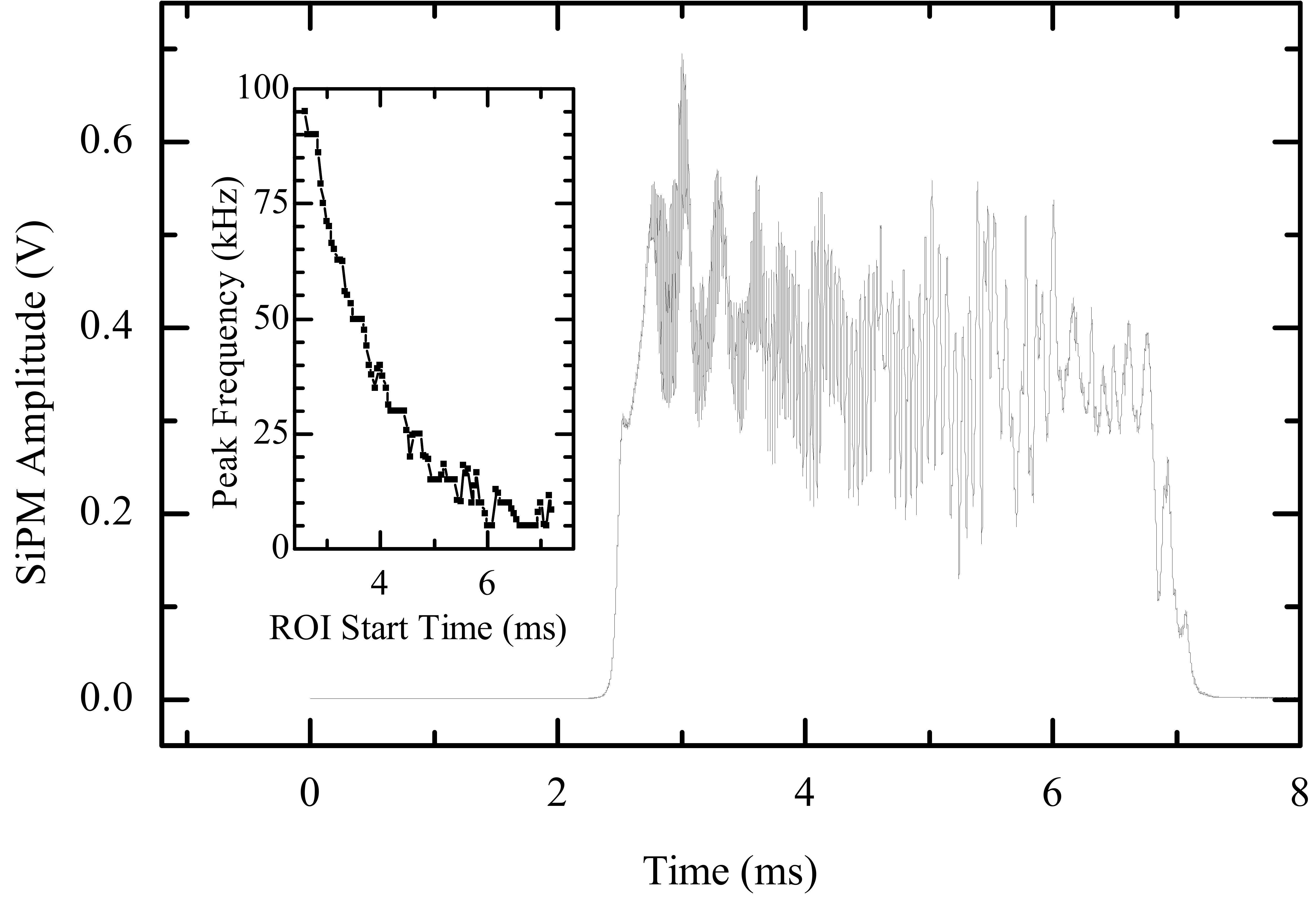}
    \caption{Full extraction trace for a cold dense plasma with a MCP gain of $\sim 2000$. Following the fast rise used for the temperature diagnostic, the slow oscillations are evidence of plasma oscillations.  The inset plots the peak FFT frequency using a rolling region-of-interest (ROI) of width $0.2\,\mathrm{ms}$.  These high frequencies may be related to Kelvin-Helmholtz instabilities. In accordance with the data in the inset, the frequency of such oscillations should decline as the plasma charge decreases.}
    \label{fig:cat}
\end{figure}

With the FC, the MCP must be run at its highest gain.  While the MCP is linear for the first electrons that escape, it soon saturates at this gain.  With the SiPM, however, the SNR is sufficiently high that, for most plasmas, the MCP can be run at less than its maximum gain.  The gain can be set such that the MCP never saturates during the extraction process (see Fig.~\ref{fig:cat}). Under these conditions, several additional plasma parameters can be measured simultaneously with the temperature measurement:
\begin{enumerate}
\item From the integrated SiPM signal, we can determine a number proportional to the total plasma charge.
\item For a dense, cold plasma, the plasma's self consistent potential energy is large compared to the kinetic energy of its constituents.  Consequently, the first appearance of escaped electrons marks when the confinement barrier height is approximately equal to the plasma self-potential.  Similarly, the last appearing electron marks when the confinement barrier has flattened.  Thus, from the energy width of the escape curve, we can determine the plasma self-potential.
\item Plasma imaging\cite{povi:15} is normally done with a fast extraction lasting less than $1\,\mu\mathrm{s}$.  The extractions used to measure the plasma temperature are relatively slow because we need to measure the time-history of the extraction.  This leads to diocotron\cite{levy:65,fine:98} and possibly Kelvin-Helmholtz\cite{peur:93b} instabilities that perturb the late extraction process, signs of which can be seen in Fig.~\ref{fig:cat}.  Nonetheless, we can image the plasma during the temperature extraction process.  Even with a slow extraction, the number of plasma charges, the plasma size, and the plasma radial position can often be inferred from the plasma image with $\sim 10\%$ precision despite the instabilities, particularly if the plasma density is low or if the temperature is high.
\end{enumerate}

\section{Conclusion}
A SiPM-based diagnostic has numerous advantages over a FC-based diagnostic for measuring the temperature of a plasma confined in a Penning-Malmberg trap. The SiPM device features enhanced robustness with low-voltage operation at high speed. It permits measurements with the sensitivity of a photomultiplier tube in an environment where a photomultiplier cannot be used, and can achieve single plasma electron resolution when coupled with an MCP and phosphor screen.

Much of the advantage of a SiPM over a FC comes from the reduced noise of the SiPM-based system.  It is somewhat surprising that the SiPM system, which counts electrons by converting the electrons to light, and then converting them back into electrons, is quieter than the FC system which counts electrons directly. This is particularly surprising because most of the light is lost; only $\sim 0.001\,\mathrm{sr}$ of the light is collected by the SiPM. Some of this loss is compensated for by the gain of the phosphor screen.  The efficiency of the P47 phosphors used in our screen is not well documented, but one can estimate that $10\textrm{--}100$ photons per incident electron are generated when the screen is at $4.5\,\mathrm{kV}$.\cite{hoes:01}  Whatever the SiPM-based system efficiency, it is high enough to produce much quieter signals than the microphonic-degraded FC system.

We have used the SiPM-based diagnostic to measure the temperatures of plasmas down to $10\,\mathrm{K}$, and have not obviously attained the lowest instrumental limit.  These are the lowest lepton temperatures that have been reliably measured.  Previous notable low lepton temperature measurements\cite{beck:96} required use of the universal function below $500\,\mathrm{K}$, and reported temperatures down to only $30\,\mathrm{K}$.  (Antiproton plasma temperatures of $10\,\mathrm{K}$ and below have been measured\cite{andr:10,gabr:11} using annihilation detectors. As with the SiPM, these detectors have very little noise.)  We have also been able to measure the temperature of plasmas with as few as $300$ trapped electrons.  For fewer electrons, collisions do not necessarily adequately Maxwellianize the plasmas, particularly as the collision rate can be strongly suppressed by O'Neil's adiabatic invariant.\cite{glin:92}

\begin{acknowledgements}
We thank Jonathan Wurtele for his help measuring the light gain of the CPC, and Adrianne Zhong for discussions about the temperature analysis. This work was supported by the DOE OFES and NSF-DOE Program in Basic Plasma Science.

The data that support the findings of this study are available from the corresponding author upon reasonable request.
\end{acknowledgements}



%

\end{document}